\documentclass{eas}
\usepackage{graphicx}
\pdfoutput=1
\usepackage{amssymb,amsmath}

\newcommand{\aap}{A\&A}  
\newcommand{\aj}{AJ}      
\newcommand{\apj}{ApJ}      
\newcommand{\apjl}{ApJL}   
\newcommand{\mnras}{MNRAS} 
\newcommand{\pasj}{PASJ}   

\newcommand{\be}{\begin{equation}}
\newcommand{\ee}{\end{equation}}
\newcommand{\bea}{\begin{eqnarray}}
\newcommand{\eea}{\end{eqnarray}}
\newcommand{\bc}{\begin{center}}
\newcommand{\ec}{\end{center}}

\newcommand{\msun}{{\rm M}_{\odot}}
\newcommand{\rsun}{{\rm R}_{\odot}}
\newcommand{\lsun}{{\rm L}_{\odot}}

\runningtitle{Mohamed {\em et al.\/}: 3D Models of Betelgeuse's Bow Shock}
\begin{document}

\title{3D Smoothed Particle Hydrodynamics Models of Betelgeuse's Bow Shock} 
\author{Shazrene Mohamed}\address{South African Astronomical Observatory, P.O. Box 9, 7935 Observatory, South Africa}
\author{Jonathan Mackey}\address{Argelander Institut f\"{u}r Astronomie, Auf dem H\"{u}gel 71, Bonn D-53121, Germany}
\author{Norbert Langer}\sameaddress{2}
\begin{abstract}
Betelgeuse, the bright red supergiant (RSG) in Orion, is a runaway star. 
Its supersonic motion through the interstellar medium has resulted in the formation 
of a bow shock, a cometary structure pointing in the direction of motion. We present the 
first 3D hydrodynamic simulations of the formation and evolution of 
Betelgeuse's bow shock. We show that the bow shock morphology 
depends substantially on the growth timescale for Rayleigh-Taylor versus  
Kelvin-Helmholtz instabilities.  We discuss our models in light of the recent {\it Herschel}, 
{\it GALEX} and {\it VLA} observations. If the mass in the bow shock shell is low ($\sim$few$\times10^{-3}\,\msun$), 
as seems to be implied by the {\it AKARI} and {\it Herschel} observations, 
then Betelgeuse's bow shock is very young and is unlikely to have reached a steady state.  
The circular, smooth bow shock shell is consistent with this conclusion. We further discuss the 
implications of our results, in particular, the possibility that Betelgeuse may have only recently 
entered the RSG phase. 
\end{abstract}
\maketitle
\section{Introduction}

From the shoulder of Orion `The Hunter' (Greek mythology), 
to stories about the fierce, red lion (Southern African mythology), 
Betelgeuse has long been a prominent part of the night sky. 
As the nearest and brightest star of its kind, Betelgeuse is now 
considered the prototype red supergiant (RSG). Estimates of its mass range 
from 8\,$\msun$ - 20\,$\msun$ and although it is very cool 
(T$_{\rm eff}$$\sim$3\,300\,K), it is highly luminous ($L_*\sim$10$^5$\,$\lsun$) 
due to its large stellar radius ($R_*\sim$1000\,$\rsun$) (see for example, 
Smith {\em et al.\/} \cite{Smi09}; Neilson {\em et al.\/} \cite{Hil11}). Its tenuous 
atmosphere is only loosely bound, consequently it loses 
$\sim2-4\times10^{-6}\,\msun$yr$^{-1}$ via a slow, $\sim$17\,km\,s$^{-1}$ wind 
(Noriega-Crespo {\em et al.\/} \cite{Nor97}, Bernat {\em et al.\/} \cite{Ber79}). 
The mechanism by which this material is lost is still unclear, but 
the process has occurred for thousands of years forming an extensive 
circumstellar envelope (CSE) of gas and dust.

Noriega-Crespo {\em et al.\/} (\cite{Nor97})  successfully imaged the CSE 
at 60 and 100 $\mu$m using {\it IRAS}. They detected  a bow shock arc $\sim$6 arcmin 
in radius to the north-east of Betelgeuse and a mysterious 
linear `bar-like' structure, located just ahead of the arc. A decade later, Ueta {\em et al.\/} 
(\cite{Ueta08}) confirmed the detection, imaging part of the bow shock arc and bar in the 
far-infrared with {\it AKARI}. More recently, high resolution {\it Herschel} observations 
revealed that the bow shock consists of multiple arcs (Cox {\em et al.\/} \cite{Cox12}, Decin {\em et al.\/} 
\cite{Dec12}). Le Bertre {\em et al.\/} (\cite{Bet12}) identified a faint, {\it GALEX} far-ultraviolet  arc 
at the same position as the outermost {\it Hershel} one.  The bar, however, was 
not detected at these shorter wavelengths. With the {\it VLA}, they also found atomic hydrogen 
coincident with the bow shock and an inner, cometary shaped detached shell of HI emission   
$\sim$4 arcmin in diameter.

Betelgeuse is moving supersonically relative to the local interstellar medium (ISM)\footnote{Although 
Betelgeuse is the only RSG with a bow shock, theoretical models predict that up to 30\% of RSGs can be runaway stars (Eldrige {\em et al.\/} \cite{Eld11}). The origin of the Betelgeuse's high space velocity is unclear but may be due to a dynamical ejection from a cluster and/or a supernova kick.},  
and its bow shock is formed by the collision of its stellar wind with this medium. 
Assuming it has reached a steady state, the bow shock can be used to probe 
the physical properties of these interacting  flows. The bow shock `radius', known as the stand-off distance, $R_{\rm SO}$, is
 the location where the ram pressures of the ISM and stellar wind are in equilibrium, and is given by: 
 \be
\rho_{\rm ISM} v^2_* =  \dot{M}_{\rm w} v_{\rm w} / 4 \pi R_{\rm SO}^2 \,, 
\label{eq: ram}
\ee
(assuming a spherical wind) where  $\dot{M}_{\rm w}$ is the wind mass-loss rate, $\rho_{\rm ISM}$ and $\rho_{\rm w}$ 
are the density of the ISM and stellar wind, respectively; $v_*$ is the velocity of the star 
with respect to the ISM, and $v_{\rm w}$ is the stellar wind velocity. Assuming momentum conservation and that the stellar wind and ISM mix 
and cool instantaneously (the thin-shell approximation), the shape of the bow 
shock is given by:  
\be
R(\theta) = R_{\rm SO}\, \mathrm{cosec}\, \theta \sqrt{3 (1 - \theta\cot \theta)}\,,
\label{eq: shape}
\ee
where $\theta$ is the polar angle measured from the axis of symmetry (Wilkin \cite{Wil96}).

Utilising these  analytic models and current estimates for Betelgeuse's wind 
 and distance, Ueta {\em et al.\/} (\cite{Ueta08}) derived a space 
velocity of $v_*$ = 40 n$_{\rm H}^{-1/2}$ km\,s$^{-1}$ with respect to the local ISM.  
Estimates of the ISM density, n$_{\rm H}$, range from 0.3 cm$^{-3}$ to 1.5 - 1.9 cm$^{-3}$, thus 
Betelgeuse's space velocity is likely to be between 73 km\,s$^{-1}$ and 28 km\,s$^{-1}$, respectively.
 Mohamed {\em et al.\/} (\cite{Moh11}) simulated models for these parameters 
 and compared the results to the {\it IRAS} and {\it AKARI} observations. 
In this paper, we highlight the main points of that study and discuss the 
conclusions in light of the recent {\it Herschel}, {\it GALEX} and {\it VLA} 
observations. 

\section{Model}

The bow shock is modeled in 3D using Smoothed Particle Hydrodynamics 
(SPH), a Lagrangian method particularly suited to studying hydrodynamical 
flows with arbitrary geometries. Throughout this study we use the {\small 
GADGET-2} SPH code (\cite{Spr05}) which we have modified to include  
stellar winds (Mohamed \& Podsiadlowski \cite{Moh07}), an ISM flow 
(Mohamed \cite{Moh10}), and atomic and molecular radiative cooling 
(Smith \& Rosen \cite{Smi03}). 

\begin{figure}[tp!]
\centering
\includegraphics[scale=.4, angle=0,trim= 0 60 0 100]{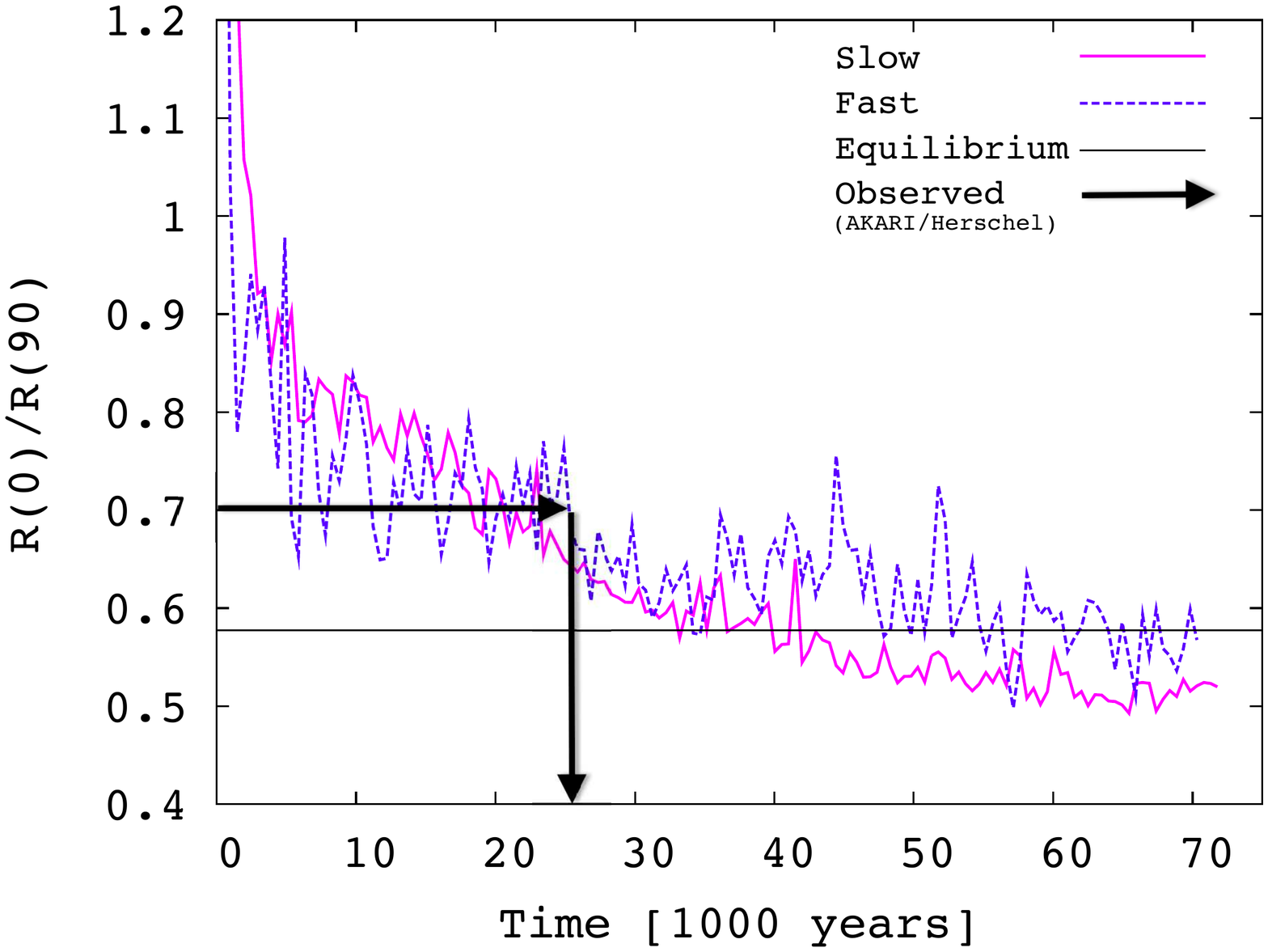}
\caption{Evolution of the ratio of $R(0^\circ)/R(90^\circ)$  for 
 a slow, $v_* = 32$\,km\,s$^{-1}$ model (magenta), and fast, 
 $v_* = 73$\,km\,s$^{-1}$ model (dashed blue), compared to the analytic, 
 steady state value, $1/\sqrt{3}$ (solid black line). The observed 
 {\it Herschel}/{\it AKARI} ratio is $\sim$0.7 (arrows).  
\label{fig: shape}}
\end{figure}

For numerical convenience, we select the stellar rest frame with the star 
located at the origin of a rectangular box ($x,y,z$ = 0, 0, 0). Given the 
uncertainty in Betelgeuse's mass-loss mechanism, we do not model the 
wind acceleration in detail. Instead wind particles are injected isotropically 
at a radius, $R_{\rm inner}$$\sim$$10^{15}$\,cm, with velocity, 
$v_{\rm w}$$\sim$17\,km\,s$^{-1}$, and  temperature 
$T_{\rm w}$$\sim$1\,000 K. The result is a smooth, constant outflow of 
material at a rate of $3.1 \times 10^{-6}\, \msun$yr$^{-1}$. 
The ISM is also assumed to be homogeneous and flows past the star in 
the direction of the $x$ axis, interacting with the stellar wind as it does so. 
We model a range of ISM densities, n$_{\rm H}$ = 0.3, 1.0, 1.5, and 1.9 cm$^{-3}$, 
with corresponding stellar velocities, $v_*$ = 73, 40, 32, and 29 km\,s$^{-1}$, 
respectively. These number densities lie at the boundary between typical 
values expected for either a warm or cold neutral ISM, so we assume 
temperatures based on the phase diagram of the standard 
model of Wolfire {\em et al.\/}  (\cite{Wolf95}), e.g., their Fig.~3d. The temperatures, 
T$_{\rm ISM}$, are 8\,000, 1\,600, 1\,000, and 650 K, respectively. 
Additional models are also run to investigate the effect of varying the ISM 
temperature, degree of cooling and numerical resolution.

The numerical method and model set up were tested with an adiabatic model. The results 
were consistent with both theoretical expectations and previous studies 
(e.g., Wilkin \cite{Wil96}, Brighenti \& D'Ercole \cite{Bri95}). 
  
\section{Results}

The simulations begin at the start of the RSG phase. 
As the stellar wind collides with the ISM, material accumulates 
at the contact discontinuity, where part of the kinetic energy of 
the gas is thermalised. The heated ISM and stellar wind expand 
outwards from either side of the contact discontinuity; the former 
pushes into the ISM, this is known as  the forward shock, and 
the latter pushes into the stellar wind, known as the reverse shock. 
Although the stellar outflow is initially spherical, it becomes 
increasingly parabolic as the star moves through the ISM. 
Eventually a steady state is achieved at which point the global 
morphology is described by Eq.~\ref{eq: shape}. From the models 
we derive the ratio of the bow shock radius at angles $\theta=0^\circ$ and  
$\theta=90^\circ$, $R(0^\circ)/R(90^\circ)$, as a function of time (shown in 
Fig.~\ref{fig: shape}). It takes several dynamical timescales for the 
bow shock to achieve the equilibrium value $R(0^\circ)/R(90^\circ)$=$1/\sqrt{3}$. 

\begin{figure}[tp!]
\centering
\includegraphics[scale=.22, angle=0,trim= 40 50 0 40]{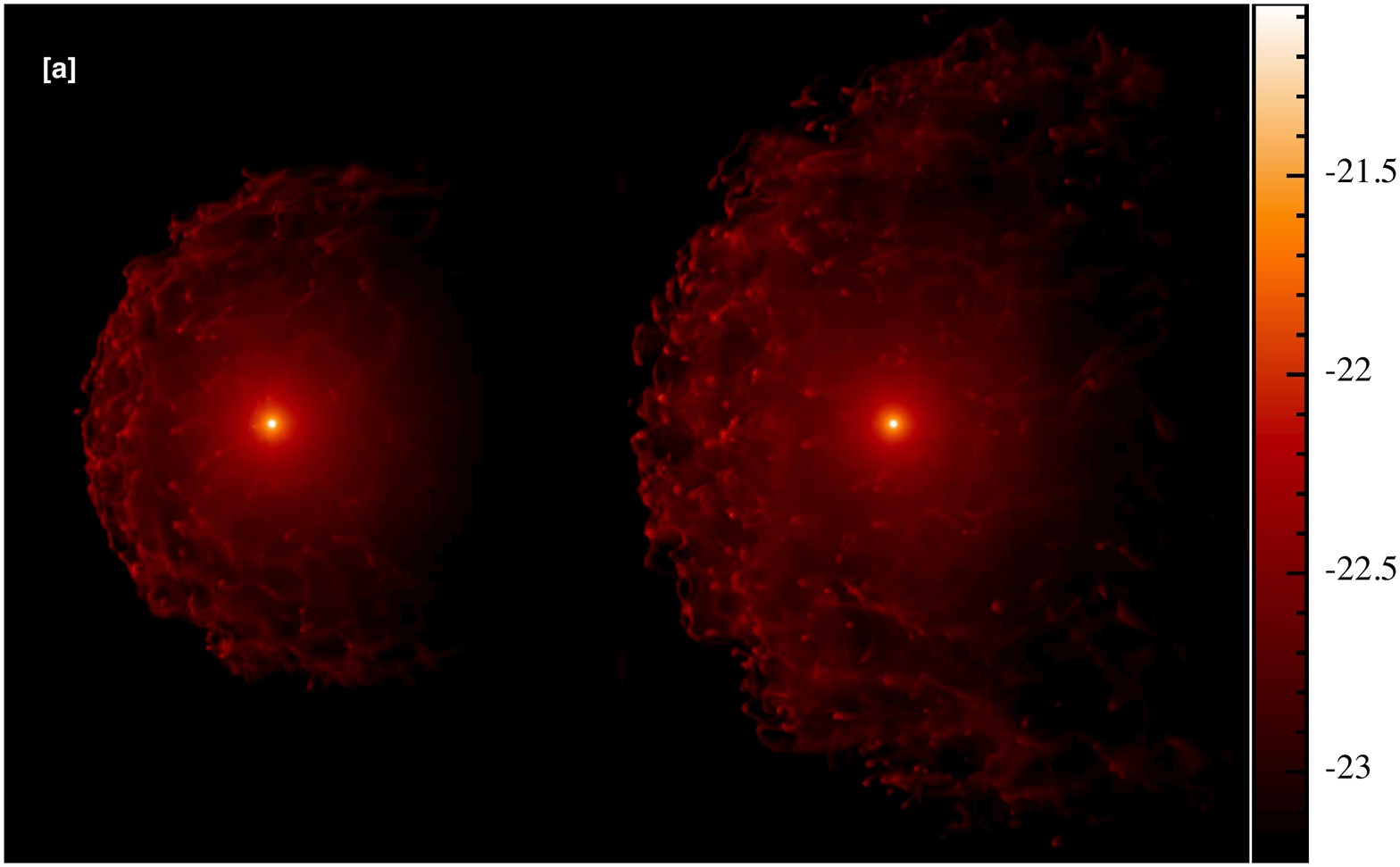}
\includegraphics[scale=.22, angle=0,trim= 40 50 60 40]{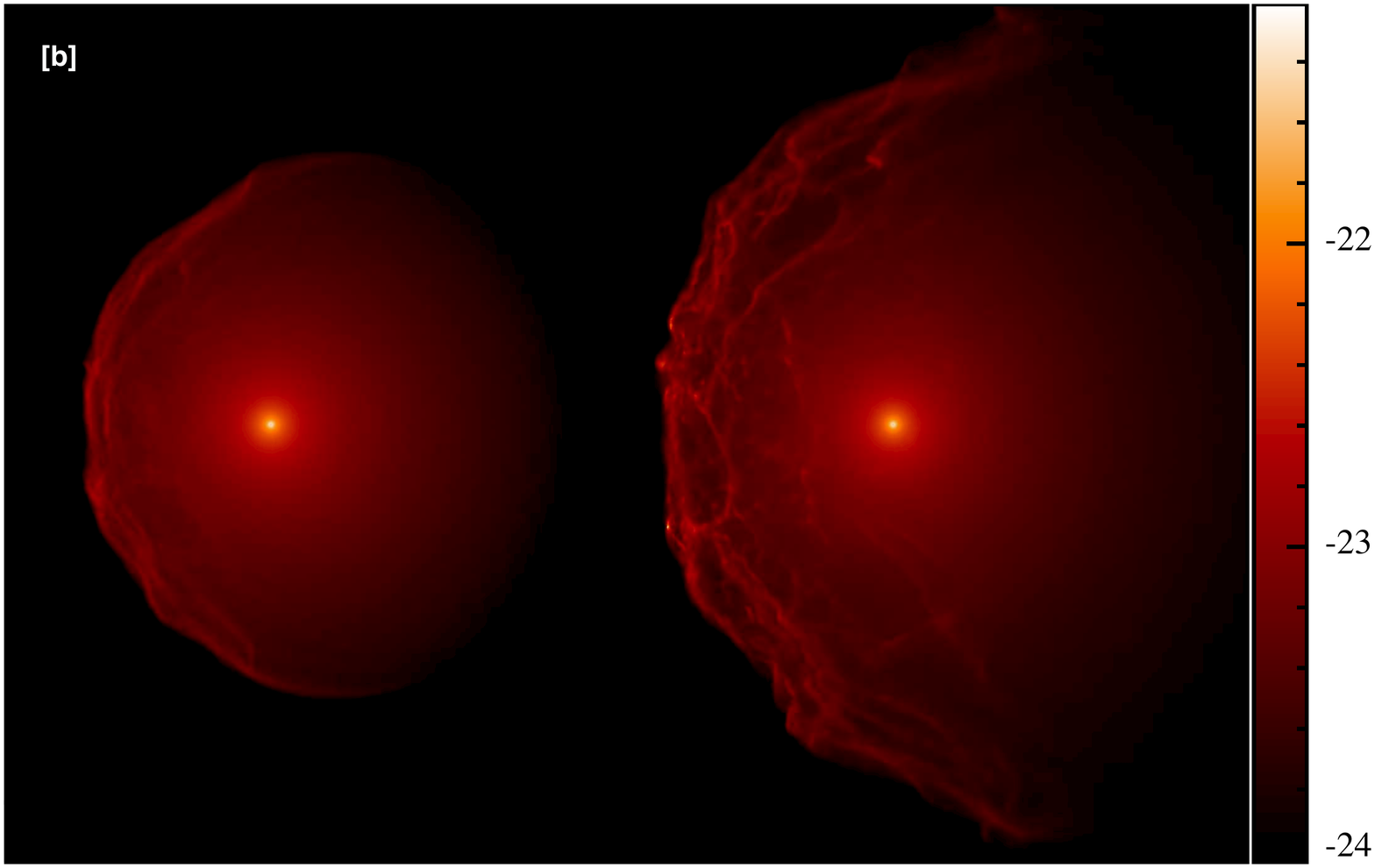}
\caption{Hydrogen column density for the slowest (29 km\,s$^{-1}$) 
[a] and the fastest (73 km\,s$^{-1}$) [b] models after 20\,000 years 
(left) and 32\,000 years (right), respectively. For animations of the 
bow shock evolution, see Mohamed {\em et al.\/} \cite{Moh11}.
\label{fig: shock}}
\end{figure}

\begin{figure}
\centering
\includegraphics[scale=.34, angle=0,trim= 50 0 235 480, clip=true]{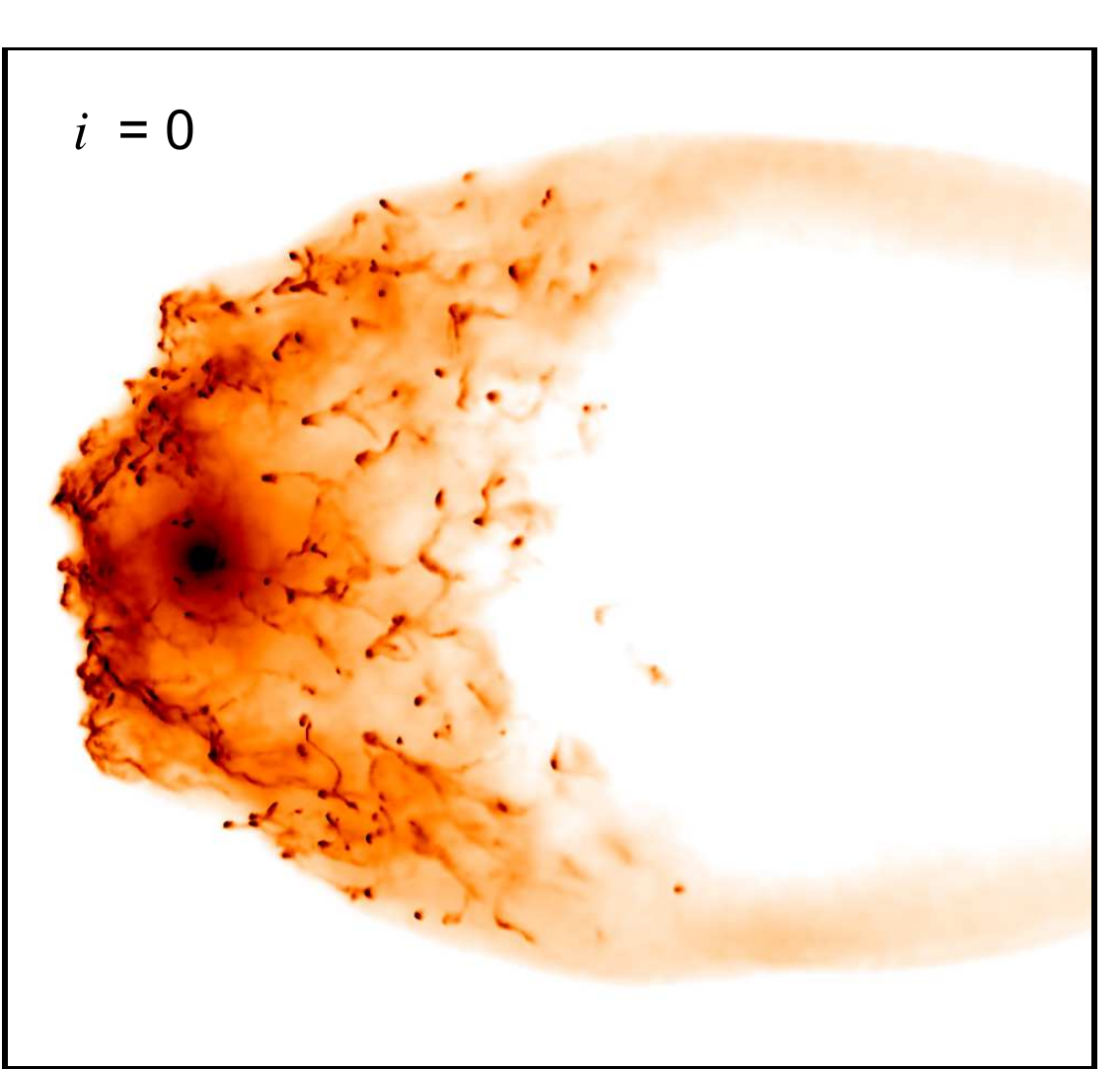}
\includegraphics[scale=.34, angle=0,trim= 45 0 235 480, clip=true]{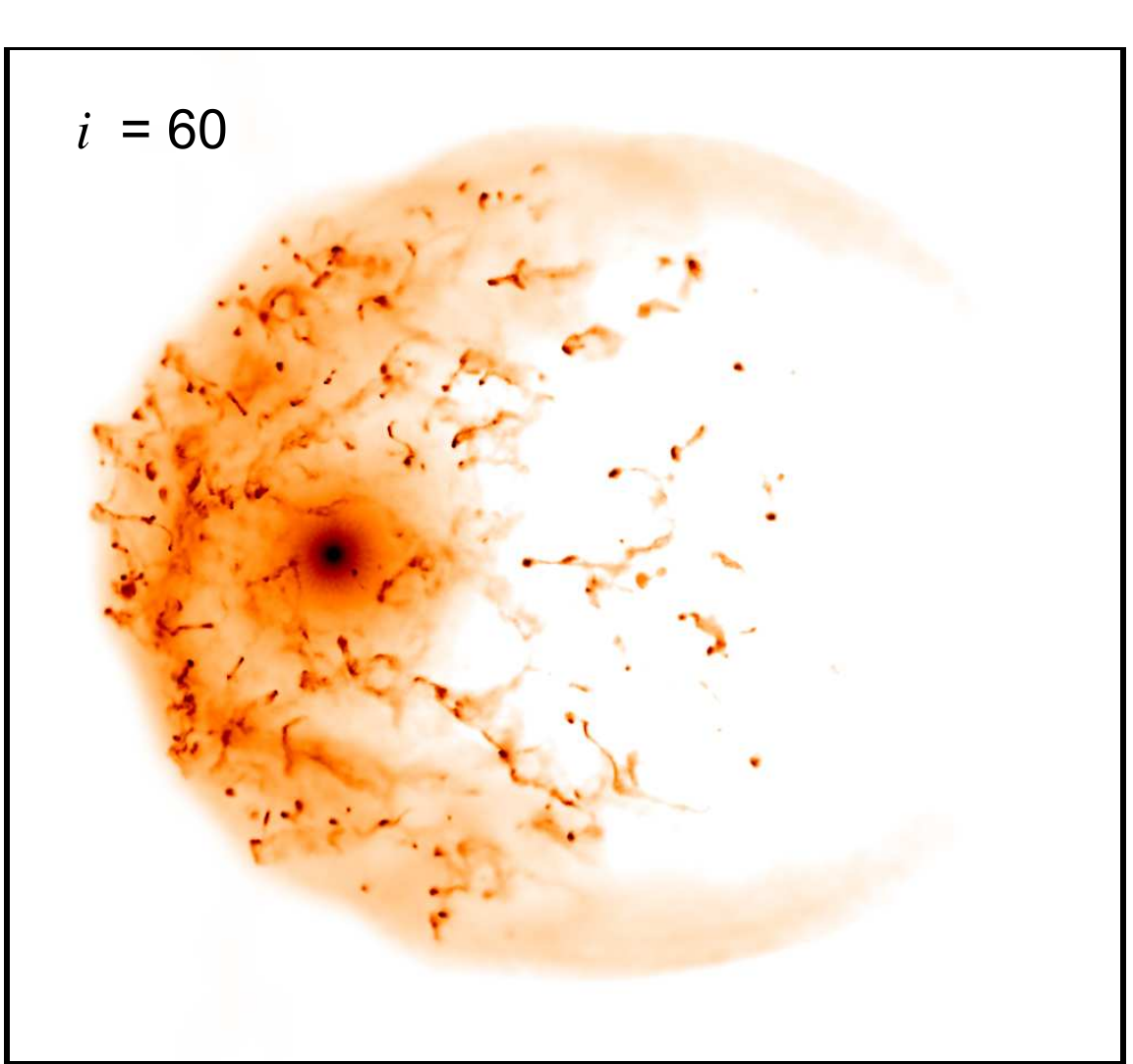}
\includegraphics[scale=.34, angle=0,trim= 42 0 235 480,clip=true]{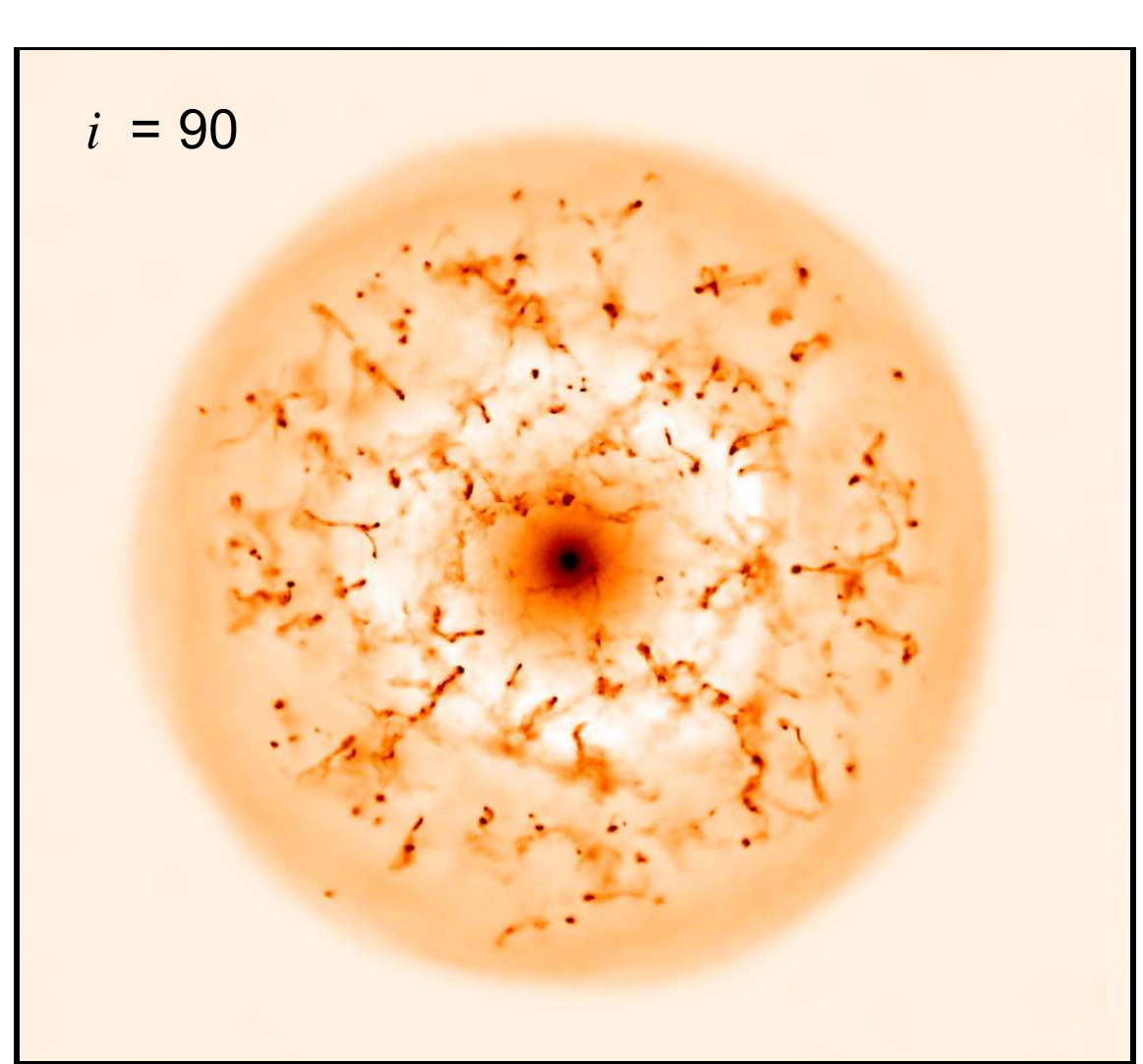}
\includegraphics[scale=.33, angle=0,trim= 45 0 235 480,clip=true]{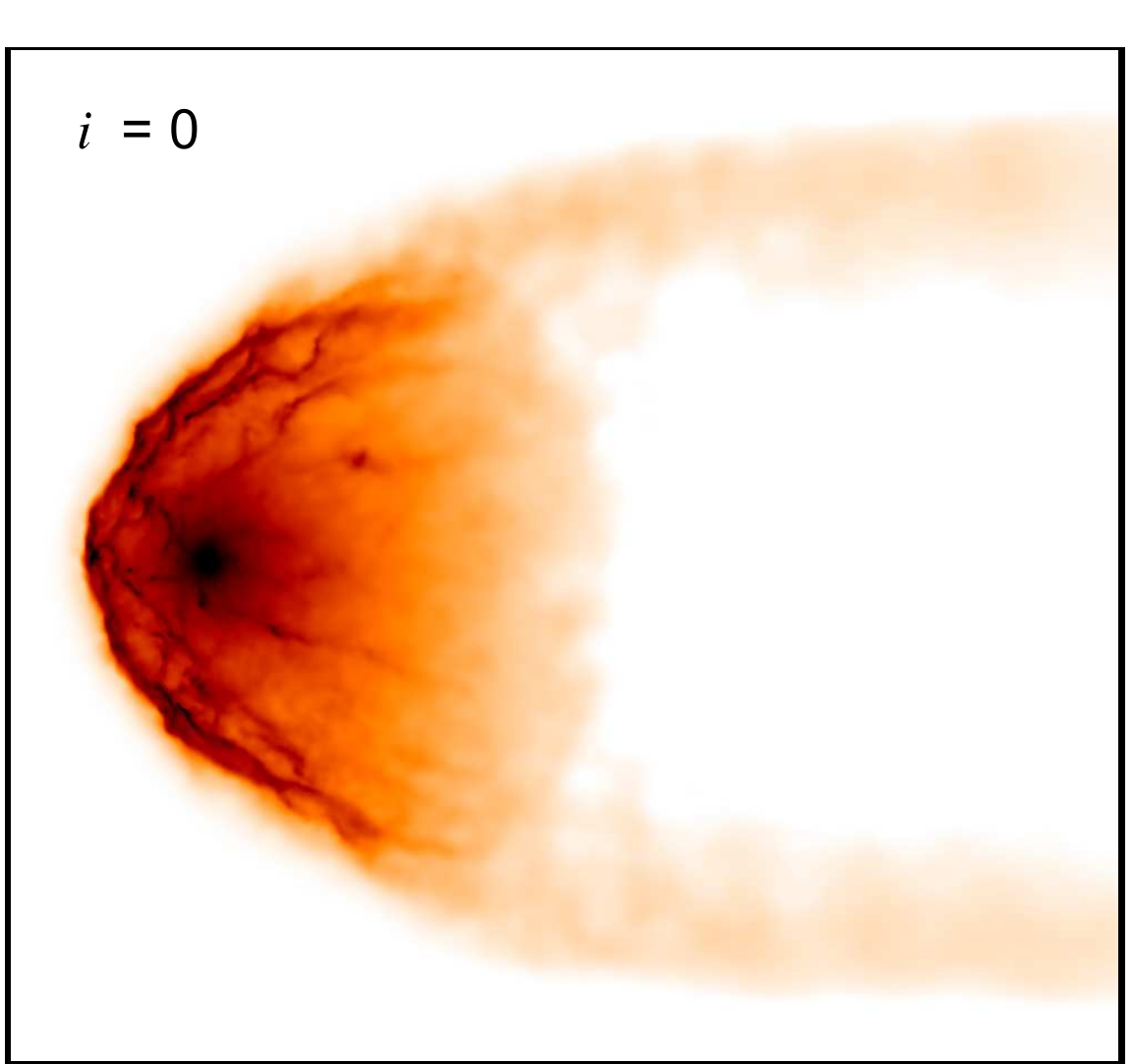}
\includegraphics[scale=.33, angle=0,trim= 40 0 230 480, clip=true]{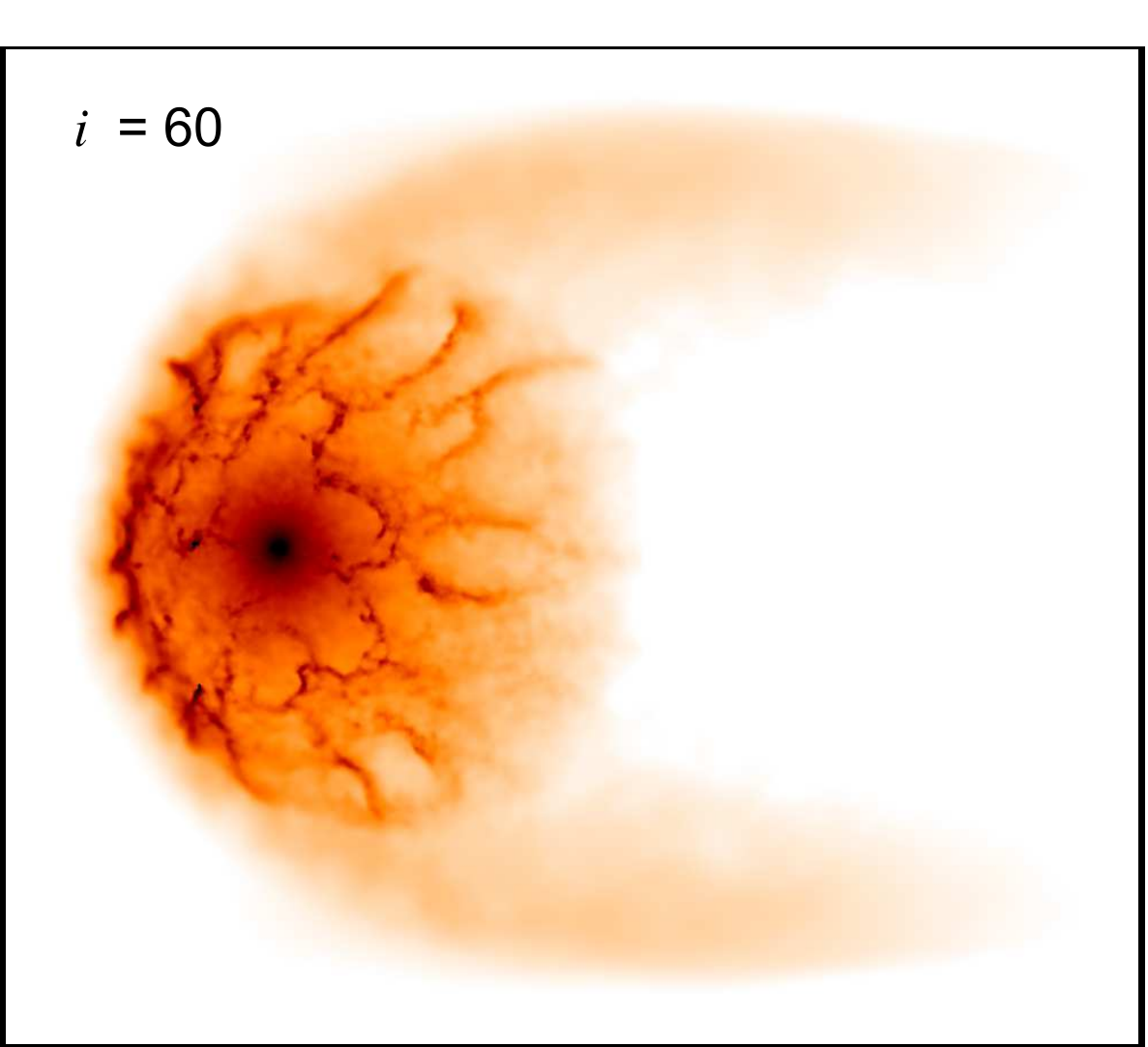}
\includegraphics[scale=.33, angle=0,trim= 40 0 230 480,clip=true]{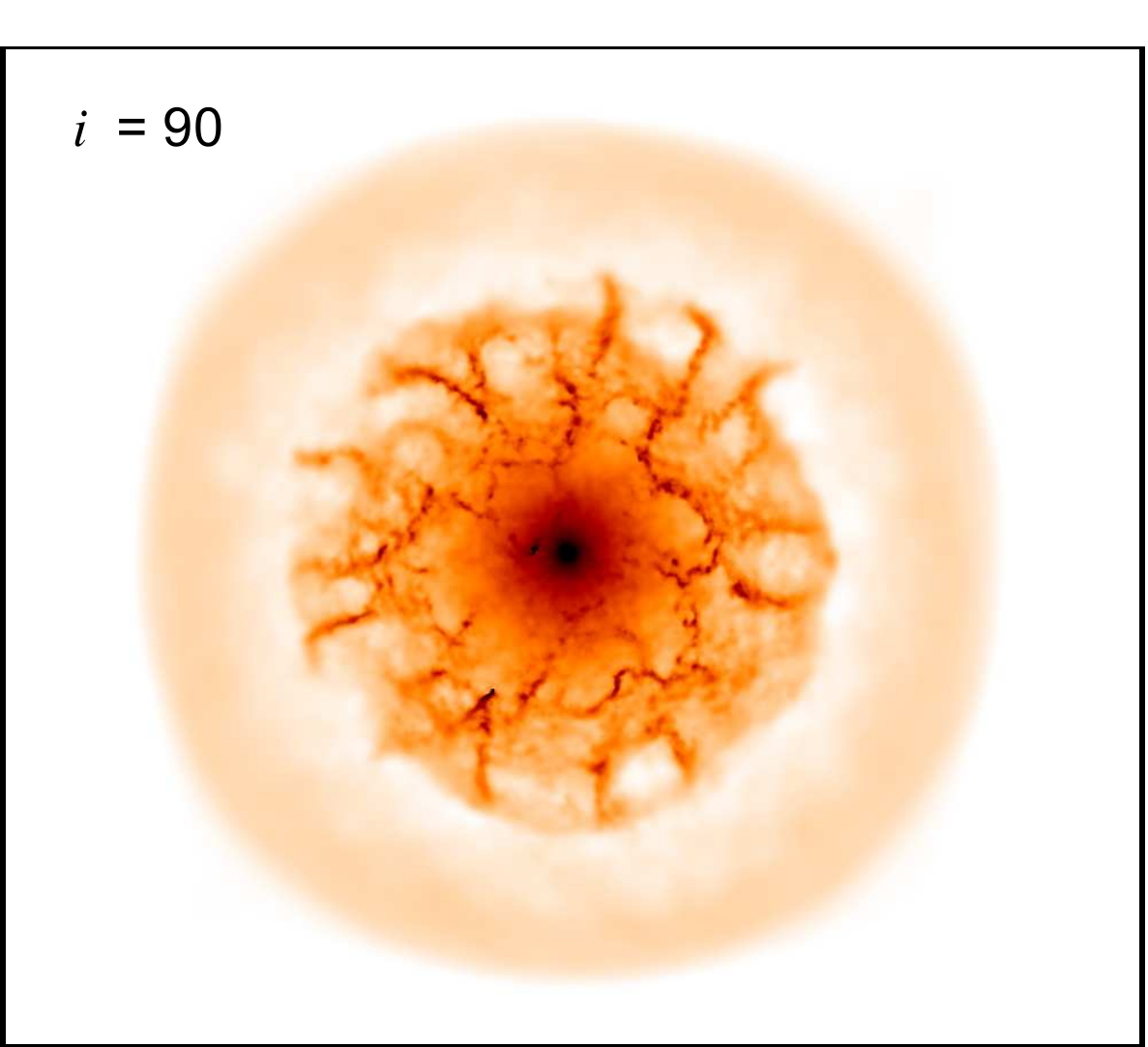}
\caption{Hydrogen column density (on a logarithmic scale) after 76\,000 
years for the slow, 32 km\,s$^{-1}$[top] and fast, 73 km\,s$^{-1}$ [bottom] 
models seen at different inclination angles, $i$. Increasing 
the inclination angle reduces the density contrast between the peak 
at the apex of the bow shock and the rest of the cometary structure, making 
the detection and identification of highly inclined systems more difficult.
\label{fig: rot}}
\end{figure}

Although all the models exhibit a similar global structure, the flow 
characteristics on smaller scales differ considerably due to the 
growth of Rayleigh-Taylor (R-T) and Kelvin-Helmholtz (K-H) 
instabilities\footnote{Rayleigh-Taylor instabilities are `finger-like' 
protrusions that occur when a light fluid is accelerated into a 
denser fluid. Kelvin-Helmholtz `rolls' or `eyes' are excited by the 
shear produced in the relative motion of two adjacent fluid layers.} 
(see Fig.~\ref{fig: shock}). In the `slow' models, $n_{\rm H}\gtrsim$1\,cm$^{-3}$ 
and $v_*\lesssim$40\,km\,s$^{-1}$, the strong cooling reduces the 
thermal pressure of the gas enabling further compression in  
the forward and reverse shocks. The greater post-shock densities reduce 
the growth timescale for R-T instabilities. This, along with the slow  space  
motion producing less shear, causes  the R-T `fingers' to develop faster 
than the K-H `rolls'. These bow shocks consist of a thin, smooth outer 
shock and a contact discontinuity that is highly distorted by R-T `fingers'. 
In the column density plots, the small-scale R-T instabilities result in 
a clumpy, knot-like sub-structure that becomes one of the dominant 
features of the bow shock, particularly when viewed at large 
inclination angles (see Fig.~\ref{fig: rot} [top]).

By contrast, the `fast' model, with $n_{\rm H}$ = 0.3\,cm$^{-3}$ and $v_*$ = 73\,km\,s$^{-1}$, is 
dominated by K-H instabilities; the greater stellar motion increases 
the shear between the ISM and stellar wind, reducing the K-H growth 
timescale. The initially lower ISM density results in less cooling and thus less 
compression; it takes much longer to grow R-T instabilities and any 
excitations that do develop are quickly advected 
downstream.  The `gentle' fluctuations of the K-H instability 
results in a more layered, filamentary appearance (see Fig.~\ref{fig: rot} [bottom]).

The appearance of the bow shock is also dependent on the emitting 
species (see Figs.~13 and 14, Mohamed {\em et al.\/} \cite{Moh11}).  
In our models, the total emissivity is a sum of the contributions from 15 different 
coolants, e.g., rotational and vibrational transitions of H$_2$, CO and H$_2$O,
 H$_2$ dissociative cooling and reformation heating, gas-grain cooling/heating, 
and atomic lines. While some species radiate from the entire bow shock surface,
e.g., H$_2$O, others are almost entirely confined to the reverse shock or forward 
shock, e.g., CO. Emission primarily from a forward shock (hotter gas) results in a 
much smoother bow shock shell, e.g., the atomic line radiation, whereas 
emission from the reverse shock  produces a more layered 
structure, e.g., collisional excitation of H$_2$O with  H$_2$. 
Several coolants, such as gas-grain, rotational transitions of CO and H$_2$O, and the
heating species produce a more `finger-like', clumpy bow shock sub-structure. 

\begin{figure}[t!]
\centering
\includegraphics[scale=.36, angle=0,trim= 40 0 70 80, clip=false]{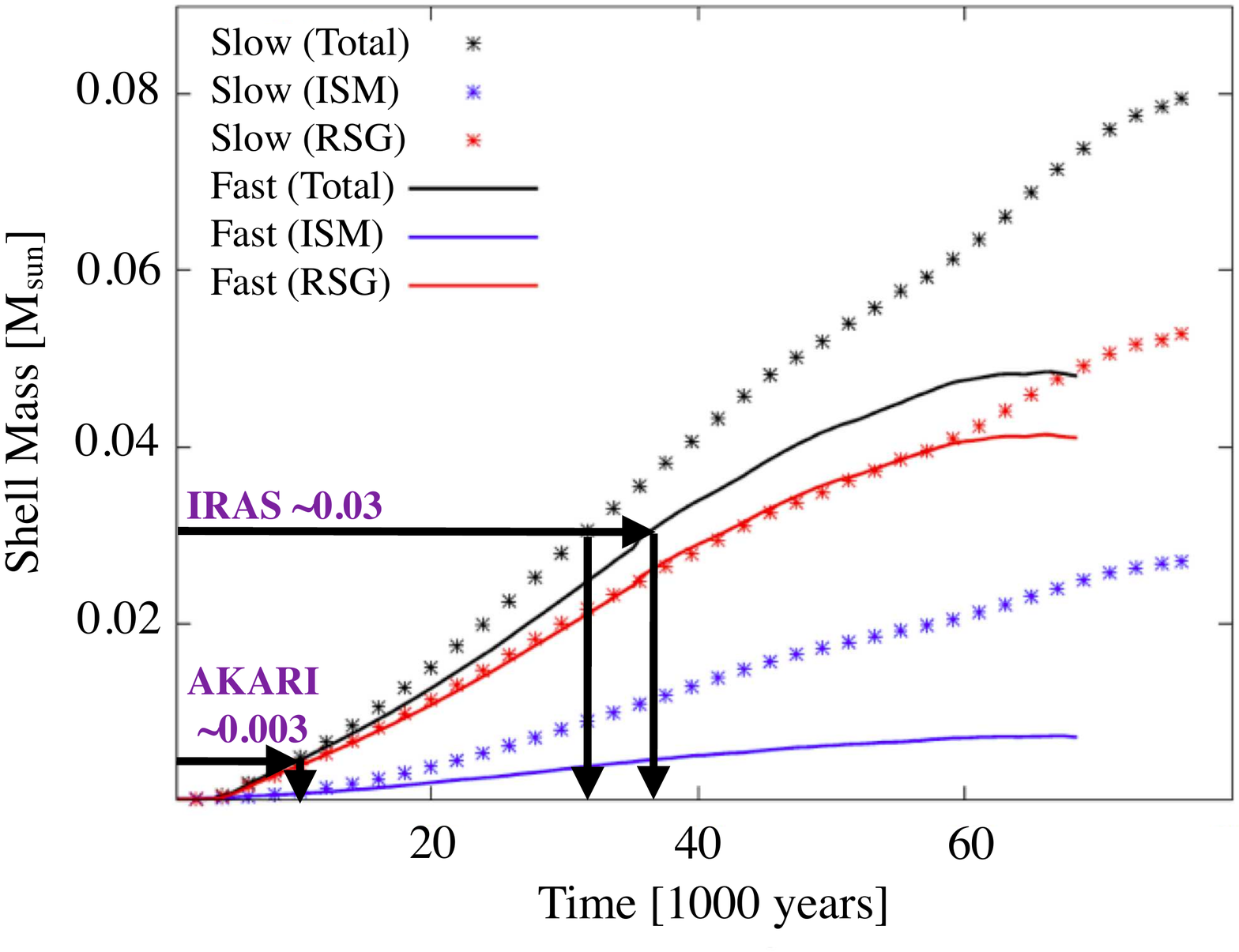}
\caption{The minimum bow shock shell mass as a function of time for a slow model (points)
and the fast model (lines). For both models, the RSG and ISM contributions are 
in red and blue, respectively, and their combined mass is plotted in black.   
The masses derived from the {\it IRAS} and {\it AKARI} observations are 
$\sim$0.03$\msun$ and $\sim$0.003$\msun$, respectively (indicated by arrows).
\label{fig: shellmass}}
\end{figure}

\section{Discussion}

Assuming no other energy sources are present and that during the collision 
all the kinetic energy of both the ISM and stellar wind is thermalised, the 
theoretical upper limit for the bolometric luminosity of the bow shock is given by 
$\dot{E}_{\rm tot} \approx \frac{1}{2}\dot{M}_{\rm w} v_{\rm w}^2 
+ \frac{1}{2}\dot{M}_{\rm w} v_{\rm *}^2$. For the parameters adopted  
in our models, $\dot{E}_{\rm tot}$ does not exceed 
$\sim$$6\times 10^{33}$\,ergs\,s$^{-1}$. In reality, however, only a small 
fraction of the kinetic energy is  radiated from the bow shock; $\sim$16\% 
and $\sim$29\% for the fast and slow models, respectively. The {\it AKARI} 
($65\mu$m)  and the {\it IRAS} ($60\mu$m) luminosities are $\sim$7$\times 10^{33}$ ${\rm ergs\,s}^{-1}$ 
and  $\sim$5$\times 10^{34}$ ${\rm ergs\,s}^{-1}$, respectively. The latter exceeds 
the theoretical upper limit for the bolometric luminosity by almost an order of magnitude, but is likely 
overestimated due to contamination from the bar and Betelgeuse itself, 
which is very luminous in the infrared. Whereas the more recent {\it Herschel} 
observations are in good agreement with the {\it AKARI} values 
(Decin {\em et al.\/} \cite{Dec12}). Although the luminosities based on 
 these higher resolution observations are consistent with the theoretical 
upper limit, as discussed above, only a small fraction 
of the kinetic energy is thermalised. Furthermore, an even smaller fraction of this 
will be radiated in the far-infrared; from our simulations the combined luminosity from 
 species thought to be responsible for the far-infrared emission, i.e.~dust 
grains, C and O fine structure lines, is at least three orders of magnitude lower than 
the observed flux. The most likely explanation is that Betelgeuse's radiation, and the 
radiation produced by hot gas in the bow shock itself, are absorbed and reemitted 
by the gas and dust in the far-infrared. 

The evolution of the bow shock shell mass is shown in Fig.~\ref{fig: shellmass}. Only 
shocked material with $x<0$ in the bow shock head is included (recall, in the models 
the star is stationary and positioned at $x, y, z$= (0,0,0) with the ISM 
moving in the direction of the $+x$ axis); this yields a lower limit for the mass in the 
bow shock at any particular time. Assuming the average mass-loss rate of the star 
has not varied significantly, we can compare the bow shock shell mass derived 
from the models with observational estimates and constrain the age of the bow shock. The bow shock mass 
derived from the {\it IRAS} flux is 0.033 $\msun$ and corresponds to a bow shock age 
of $\sim$35\,000 years (see Fig.~\ref{fig: shellmass}, arrows). However, as discussed 
above, the {\it IRAS} flux is likely an overestimate, thus the age is an upper limit. 
The {\it AKARI} and {\it Herschel}  bow shock masses are an order of 
magnitude lower, $\sim$0.0033\,$\msun$ and $\sim$0.0024\,$\msun$ 
(Decin {\em et al.\/} \cite{Dec12}), respectively, which would 
imply an age of $\sim$10\,000 years. If this is the case, however, the wind would not have had 
sufficient time to expand to the current bow shock radius. One possible solution is
that the observed shell mass is underestimated due to uncertainties in the 
flux to mass conversion (e.g., the distance to the star, the gas-to-dust ratio, dust emissivity). In our models the 
shell takes $\sim$20\,000 years to reach the observed bow shock radius 
 by which time the mass in the bow shock is 
approximately $0.02\,\msun$. This is higher than the value obtained from the far-infrared
observations, but may be consistent  within the uncertainties, and is in agreement with the 
masses based on 21cm neutral H observations (Le Bertre {\em et al.\/} \cite{Bet12}, Decin {\em et al.\/} \cite{Dec12}).  
Note, however, that at $\sim$20\,000 years none of our models are close to reaching a steady state.

The shape of Betelgeuse's bow shock is more circular than parabolic. As shown 
in Fig.~\ref{fig: rot} the bow shock becomes increasingly circular with larger inclination 
angles, i.e.~at large angles between the apex of the bow shock and the plane of the sky.  
Ueta {\em et al.\/} (\cite{Ueta08}) derived an inclination of 56$^\circ$ using Eqs.~\ref{eq: ram} 
and \ref{eq: shape} (i.e.~assuming a steady state)  which is consistent with 
$\sim$50$^\circ$ based on the tangential and radial velocities. However, an 
alternative explanation for the circular shape is that 
the bow shock has not yet reached a steady state. From the {\it Herschel} and  {\it AKARI} observations, the ratio of 
$R(0^\circ)/R(90^\circ)$ is approximately 0.7, which is much greater than 
the equilibrium value and corresponds to an age of $\lesssim$30\,000 years (Fig.~\ref{fig: shape}). 

The multiple arcs and even their bright knots in the {\it Herschel} 
observations closely resemble the filamentary structure of the fast 
model (Fig.~\ref{fig: shock}[b]). The filaments arise when the  
K-H instabilities are seen projected onto the plane of the sky. This 
projection effect could account for some of the arcs and other   
structures observed at locations well inside the bow shock 
radius (although the very large mass of the HI detached shell 
would be difficult to explain). From the radial velocity and proper 
motion, the space velocity of Betelgeuse  is unlikely to 
be as high as 73 km\,s$^{-1}$.  (Understanding the origin of the 
far-ultraviolet emission may put constraints on the upper end of 
the stellar velocity.) The similarity between the observations  
and the fast model,  and the lack of clumpy sub-structure that 
characterised the slow models, suggests that the bow shock is dominated 
by K-H rather than R-T instabilities. This situation could occur for the slow 
 models if the RSG wind expands into a much lower density, hot ISM. 
These conditions reduce the cooling and hence compression in the bow shock,  
increasing the growth time for R-T instabilities. However, for the slow model, 
such conditions are not consistent with the relation $v_*$ = 40 n$_{\rm H}^{-1/2}$ km\,s$^{-1}$, 
i.e.~the ram pressures are not in equilibrium and the bow shock is not yet in a steady state. 
Thus the overall smooth appearance and lack of well-developed instabilities  
further strengthen the argument that the bow shock must be young.

\section{Implications}

We find that many of the physical and morphological characteristics of 
Betelgeuse's bow shock, e.g., the smooth circular shape, 
the low shell mass and multiple arc sub-structure, are consistent 
with a young bow shock ($\lesssim$30\,000 years). Consequently, within 
this time frame, the local ISM through which the star is moving and/or the stellar 
wind must have undergone significant changes.  In Mohamed {\em et al.\/} 
(\cite{Moh11}) we proposed that such dramatic changes may have occurred if 
Betelgeuse only recently became a RSG, transitioning from either a main sequence 
(MS) star or a blue supergiant (BSG) (i.e.~moving from the `blue' to the `red' in the 
Hertzsprung-Russell diagram). The radius of Betelgeuse's MS or BSG wind bubble would 
have been $\sim$1 pc, assuming typical wind mass-loss rates ($\sim$$10^{-7}\,\msun\,$yr$^{-1}$) 
and wind velocities ($\sim$$10^3$\,km\,s$^{-1}$) for such hot, blue stars. 
A RSG phase of a $\sim$few$\times$10\,000 years would bring the star 
close to the edge of such a bubble; thus, the mysterious `bar' ahead of Betelgeuse's 
  bow shock could be a remnant shell produced during this earlier 
 phase of `blue' evolution (Mohamed {\em et al.\/} 
\cite{Moh11}). A blue-red transition would also mean that 
the RSG wind expands into a MS or BSG bubble filled with low density, 
hot gas, precisely the conditions required to form the observed smooth, K-H dominated, 
multiple arc characteristics of Betelgeuse's bow shock. 
 
 Mackey {\em et al.\/} (\cite{Mac12}) carried out a detailed investigation 
 of a BSG to RSG transition, including an evolving wind with a 
 non-constant mass-loss rate and wind velocity. Their models reproduce 
  the observed bow shock mass and multiple arcs. They also find that 
  the receding BSG bow shock is a plausible candidate for the linear bar 
  structure. More detailed comparisons of the models with the observations  
  will require a more sophisticated  treatment of dust, radiative transfer and possibly 
  magnetic fields (see Decin {\em et al.\/} \cite{Dec12}). 

Further observations are also required to reduce the number of 
free and uncertain parameters in the models. In particular, a more accurate 
distance to Betelgeuse (197$\pm$ 45\,pc, see Harper {\em et al.\/} \cite{Harp08}) 
would reduce the uncertainty in several key areas, e.g.,  in deriving 
the space velocity of the star (to this end a more accurate proper motion and 
radial velocity are also needed).  The bow shock mass also depends  
on the distance as well as the highly uncertain dust properties, e.g., composition, the 
dust-to-gas ratio and the dust temperature. Future observations, e.g., 
with  {\it ALMA}, could constrain the gas density, temperature and 
velocity structure in the CSE. Indeed, tracing material from the stellar photosphere all the way 
to the bow shock would give us insight into the mass-loss history 
 of Betelgeuse, a key ingredient in stellar evolution models.

\section*{Acknowledgements}The author is grateful to the organizers for financial support 
towards conference expenses. The rendered figures in this paper were made using 
the {\small SPLASH} visualization code (Price \cite{Pri07}).


\end{document}